\documentclass[nofootinbib,preprint,amsmath,amssymb,aps]{revtex4-1}

\usepackage{amsmath,amsfonts,amssymb,slashed,color,empheq,tensor}
\usepackage[svgnames]{xcolor}
\usepackage[pdftex,breaklinks,colorlinks=true,urlcolor=RoyalBlue,%
linkcolor=blue,citecolor=blue]{hyperref}

\usepackage[utf8,latin1]{inputenc}
\usepackage{graphicx}
\usepackage{dcolumn}

\usepackage{mathabx}
\usepackage{mathrsfs}  
\usepackage{cases}
\usepackage{bm}
\newcommand{\qm}[1]{``#1''}
\usepackage{hyperref}
\usepackage{stackengine,scalerel}
\hypersetup{colorlinks, linkcolor={red},citecolor={blue},urlcolor={blue}}  
\usepackage{bbold}

\def\R{{\mathbb R}} \def\C{{\mathbb C}} 
 \def\one{\mbox{1 \kern-.59em {\rm l}}}

\newcommand{\Tr}{\mathrm{Tr}}

\def\cA{{\cal A}}   
 \def\cE{{\cal E}}  
   
\def\cJ{{\cal J}} \def\cK{{\cal K}} \def\cL{{\cal L}} 
\def\cM{{\cal M}} \def\cN{{\cal N}}  
  \def\cR{{\cal R}} 
 \def\cT{{\cal T}}

 \def\g{\gamma} \def\G{\Gamma}
 \def\d{\delta} 
    \def\k{\kappa}
\def\l{\lambda}   \def\m{\mu}
    \def\r{\rho}
\def\s{\sigma}  \def\t{\tau}

\newcommand{\eq}[1]{(\ref{#1})}
\newcommand{\del}{\partial}
\def\nn{\nonumber} 
\def\obar{\overline}

\begin{document}

\title[Fermions on curved backgrounds  of matrix models]{Fermions on curved backgrounds  of matrix models}

\author{Emmanuele Battista$^{1}$\vspace{0.5cm}}\email{emmanuelebattista@gmail.com}\email{emmanuele.battista@univie.ac.at} 
\author{Harold C. Steinacker$^{1}$}
\email{harold.steinacker@univie.ac.at}

\affiliation{$1$ Faculty of Physics, University of Vienna Boltzmanngasse 5, A-1090 Vienna, Austria
}

\date{\today}

\begin{abstract}
We discuss the propagation of fermions on generic, curved branes in IKKT-type
matrix models.
The Dirac operator can be understood either in terms of a Weitzenb\"ock connection, or in terms of the Levi-Civita connection with extra torsion term. We discuss in detail the coupling of spin to the background geometry  using the JWKB approximation. Despite the absence of local Lorentz invariance in the underlying IKKT framework, our  results agree with the expectations of Einstein-Cartan theory,
and differ from general relativity only by an extra coupling to the totally antisymmetric part of the torsion.  The  case of FLRW cosmic background solutions is discussed as a special case.
\end{abstract}

\maketitle

\section{Introduction}

Reconciling gravity with quantum mechanics remains one of the outstanding problems in theoretical physics. One of the proposed approaches towards this goal is provided by the IKKT matrix model, which was introduced in the context of string theory \cite{Ishibashi:1996xs}. 
In this framework, spacetime  arises as a brane-like solution, with intrinsic quantum structure. The description of the effective metric 
in this framework is by now well understood \cite{Steinacker:2010rh,Steinacker:2020xph}. However, the coupling of fermions to such a background geometry has not yet been studied in detail. This paper is dedicated to fill this gap.

 From a formal point of view, the fermions in the IKKT model are governed by an action which is completely fixed by supersymmetry, and which is {\em not} equivalent to the coupling of fermions to gravity in general relativity. However the distinction turns out to be sub-leading and rather  subtle, and a proper assessment requires a careful analysis going beyond the level of point particles.

The relativistic description of elementary particles and extended objects in a given gravitational field has a long history. The  dynamics of a spin-1/2 fermion can be addressed by generalizing  the Dirac equation to curved spacetimes, as was first carried out  by Fock, Ivanenko, and Weyl in 1929 in the framework of general relativity (see Ref. \cite{Pollock2010} for a modern approach to this topic).  The analysis in the case of the  Einstein-Cartan theory and the related  Riemann-Cartan spacetime was performed afterwards. In this context, it is found that the spin of the fermion couples to (the totally antisymmetric part of) the contorsion, i.e., the non-Riemannian part of the connection representing the geometric counterpart of the spin \cite{Hehl1976_fundations,Gasperini-DeSabbata}. On the other hand, the classical motion of a finite-size  body endowed with  a macroscopic angular momentum (usually referred to as  \qm{spin}, despite its completely classical nature) in general relativity is ruled, in the so-called pole-dipole approximation,  by the Mathisson-Papapetrou-Dixon equations \cite{Mathisson1937a,Papapetrou1951,Dixon1970}. The main consequence brought in by the underlying spin-gravity coupling is that  the particle orbit  differs from a geodesic and its spin undergoes a precession motion. For more details about the modern applications of Mathisson-Papapetrou-Dixon equations in gravity theories we refer the reader to Ref. \cite{Putzfeld2015}  and references therein.

Due to the formal analogy between the macroscopic angular momentum of an extended object and the quantum spin of an elementary particle, a  link between the classical and quantum dynamics can be established  when  a certain  semi-classical limit is invoked. Indeed, the main features of the former can be recovered from the relativistic Dirac equation, framed either in general relativity or Einstein-Cartan theory, by exploiting either the Jeffreys-Wentzel-Kramers-Brillouin (JWKB) approximation or the Foldy-Wouthuysen approach \cite{Audretsch1981,Audretsch1981b,Rudiger1981,Hayashi1990,Cianfrani2008,Obukhov2009,Obukhov2014}. This scheme can be further enlarged by considering higher-spin fields and, in particular, it turns out that  the spin precession depends on the magnitude of the  spin vector \cite{Hayashi1992}. 

In this paper, we evaluate the propagation of a spin-1/2 particle in a generic curved background provided by the IKKT matrix model. The ensuing motion  is investigated starting from a Dirac-like action and by exploiting a semi-classical particle limit, which  is worked out by means of the JWKB approximation. 
First, we show that the fermionic action in the matrix model differs from the standard form in GR only by an extra coupling to the dilaton and to the totally anti-symmetric part of the Weitzenb\"ock connection associated with the effective frame defined by the matrix model background. Based on this action, we show that the dynamics of the fermion at the first non-trivial order of the JWKB approximation does not contradict the standard expectations of gravity theories (i.e., general relativity or Einstein-Cartan model). In fact, in the most general setting, IKKT pattern predicts that both the translation and the rotational motion of the Dirac particle  have the same form as the dynamical equations of a spin-$1/2$ fermion  in a Riemann-Cartan spacetime.  This is a non-trivial result, because  local Lorentz invariance is not manifest in the IKKT framework.

The plan of the paper is as follows. After having outlined in Sec. \ref{Sec:general-framework} the properties of the general geometric framework employed, the semi-classical Dirac-like action for fermions evolving  on a generic curved background of the IKKT matrix model is analyzed in  Sec. \ref{Sec:fermions-in-IKKT}. Then, 
the dynamics of the Dirac fermion is evaluated  in Sec. \ref{Sec:particle-limit-Dirac} by means of the JWKB approach. The particular background represented by the cosmological FLRW spacetime is considered in Sec. \ref{Sec:Cosmic-solution}. Last, we draw our conclusions in Sec. \ref{Sec:concluosions}. Supplementary material is provided in the appendices.

\emph{Notations.} We use metric signature  $(-,+,+,+)$ and units $G=c=\hbar=1$. However, for the sake of clarity, in some cases we write explicitly $\hbar$ terms. $\alpha,\beta,\dots=0,\dots,3$ and $i,j,\dots=1,2,3$ are coordinate indices, whereas $a,b,\dots=\widehat{0},\dots,\widehat{3}$    and $\widehat{a},\widehat{b},\dots = \widehat{1},\widehat{2},\widehat{3}$  are tetrad indices. The flat metric is indicated by $\eta^{ab }=\eta_{ab }={\rm diag}(-1,1,1,1)$. Round (respectively, square) brackets around tensor indices stands for the usual symmetrization (respectively, antisymmetrization) procedure, e.g., $A_{(ij)}=\frac{1}{2}(A_{ij}+A_{ji})$ (respectively, $A_{[ij]}=\frac{1}{2}(A_{ij}-A_{ji})$).

\section{The general geometric framework}
\label{Sec:general-framework}

In this section, we provide the essential details of our geometrical framework. 

We consider Yang-Mills matrix models such as the IKKT model \cite{Ishibashi:1996xs}, defined by an action of the structure
\begin{align}
S[T,\Psi] = \frac{1}{g^2}{\rm Tr}\big( [T^A,T^B][T_{A},T_{B}] 
\,\, + \overline\Psi \Gamma_A[T^A,\Psi] \big) \ .
\label{MM-action-IKKT}
\end{align}
Here the $T^A$ $ (A = 0,..., D-1)$ are hermitian matrices, and $\Psi$ are fermionic matrices described below.
We want to study the propagation of fermions on 
some given background
$\{T^A\}$  in the semi-classical regime, where the backgrounds can be described as symplectic manifolds $\cM$ embedded in target space via 
\begin{align}
   T^A:\ \ \cM \hookrightarrow \R^D \ 
   \label{background}
\end{align}
and all commutators are replaced by Poisson brackets  $[.,.] \sim i \{.,.\}$.
Moreover, we restrict ourselves for simplicity to $3+1$-dimensional branes embedded along the first 4 matrix directions labeled by 
 $a=0,..,3$, setting the remaining matrices to zero. An introduction and motivation for this framework can be found in Ref. \cite{Steinacker:2010rh,Steinacker:2019fcb}, see also e.g. \cite{Chaney:2015ktw,Steinacker:2017bhb,Hatakeyama:2019jyw,Nishimura:2019qal,Sperling:2019xar,Nishimura:2022alt,Steinacker:2021yxt,Brahma:2022dsd,Karczmarek:2022ejn,Battista-Steinacker-2022} for related work in this context.

In the semi-classical regime, the effective metric on such a background is determined by the kinetic term for fluctuations in the matrix model, which can be written as\footnote{This action applies directly to transversal fluctuations $T^A \to T^A + \Phi^A$ for $A=4,...,9$ of the background, which are interpreted as scalar fields on $\cM$. However the same metric 
$G^{\mu\nu}$ also governs tangential fluctuations $T^a \to T^a + \cA^a$ of the background, which describe gauge fields on 
$\cM$ \cite{Steinacker:2010rh}.}
  \begin{align}
  S[\phi] &\sim \ -\int\limits_\cM \!\!  d y_0 \ldots d y_3 \, \rho_M \,  \g^{\mu\nu}
   \del_\mu \phi \del_\nu \phi \ 
   = \ -\int\limits_\cM\!\! d^4 y\, \sqrt{|G_{\mu\nu}|}\,
  G^{\mu\nu}\del_\mu \phi \del_\nu \phi \ .
  \label{scalar-action-G}
\end{align}    
 Here
\begin{align}
 \gamma^{\mu\nu} = \tensor{E}{^a^\mu}\tensor{E}{^b^\nu}\eta_{ab}, \qquad 
 G^{\mu\nu} := \frac 1{\r^2} \g^{\mu\nu}, 
 \label{eff-metric}
\end{align}
define an auxiliary  and the effective metric on $\cM$, respectively, in terms of the "Poisson" frame
\begin{align}
    \tensor{E}{^a^\mu} = \{T^a,y^\mu\} \ 
    \label{frame-E-def}
\end{align}
 in local coordinates $y^\mu$.
 The conformal factor or dilaton $\r$ is defined by 
\begin{align}
\r^2 &= \r_M \sqrt{|\g^{\mu\nu}|}  \,  
 \label{rho-def}
\end{align}
where $\r_M$ is the symplectic density on $\cM$.
This motivates to define 
the effective frame $\cE^{a\mu}$ by absorbing the dilaton $\r$, 
\begin{align}
   \tensor{\cE}{^a^\mu} &= \r^{-1} \tensor{E}{^a^\mu}, \label{eff-frame-and-rho}
   \\
   G^{\mu\nu} &= \tensor{\mathcal{E}}{^{a}^\mu} \tensor{\mathcal{E}}{^{b}^\nu} \eta_{{{a}}{b}}, 
\end{align}
as well as  the inverse  frames $\tensor{\mathcal{E}}{^a_\mu}$ and $\tensor{E}{^a_\mu}$ 
through
\begin{align}
\tensor{\mathcal{E}}{^{a}_\mu} \tensor{\mathcal{E}}{_{b}^\mu} = \delta^a_b  = 
\tensor{E}{^{a}_\mu} \tensor{E}{_{b}^\mu} \ 
\end{align}
so that
\begin{align}
G_{\mu\nu} &= \eta_{{{a}}{b}} \tensor{\mathcal{E}}{^{a}_\mu} \tensor{\mathcal{E}}{^{b}_\nu} \ ,
\\
\g_{\mu\nu} &= \eta_{ab} \tensor{E}{^a_\mu}\tensor{E}{^b_\nu} \ .
\end{align}
The Weitzenb\"ock connection $\tensor{\Gamma}{_\nu_\rho^\mu}$ associated to the frame $\tensor{E}{_a ^\mu}$ is defined by the condition
\begin{align}
    0= \nabla_\nu \tensor{E}{_a ^\mu} = \partial_\nu \tensor{E}{_a ^\mu} + \tensor{\Gamma}{_\nu_\rho^\mu} \tensor{E}{_a ^\rho}.
\end{align}
This connection has a vanishing curvature, but  a non vanishing torsion and contorsion tensors, which are given by
\begin{align}
\tensor{ T}{_{\r}_{\s}^\mu} &= \tensor{ \Gamma}{_{\r}_{\s}^\mu}-\tensor{ \Gamma}{_\s_\r^\mu} \ , 
 \label{torsion-BG-explicit}
 \\
  \tensor{ K}{_{\mu}_{\nu}^{\s}}&=\frac{1}{2} \left(\tensor{ T}{_{\mu}_{\nu}^{\s}}+\tensor{T}{^\sigma_\mu_\nu} -\tensor{T}{_\nu^\sigma_\mu}\right) \ . 
 \label{CONtorsion-BG-explicit}
\end{align}
Due to the specific form \eq{frame-E-def} of the frame, their traces are given by \cite{Steinacker(2020)}
\begin{align}
 \tensor{ K}{_{\mu}_{\s}^\mu} = \tensor{ T}{_{\mu}_{\s}^\mu}
  =  \frac{2}{\rho} \del_\s \r \ . 
  \label{trace-torsion}
\end{align}
 The Levi-Civita connection $ \tensor{\G}{^{(\g)}_\mu_\nu^\s}$  for the  metric $\g^{\mu\nu}$ is
\begin{align}
\tensor{\G}{^{(\g)}_\mu_\nu^\s}
&= \frac 12 \g^{\s\r}\left(\del_\mu \g_{\r\nu} 
+ \del_\nu \g_{\r\mu}
- \del_\r \g_{\mu\nu}\right) = \tensor{\G}{_\mu_\nu^\s}  - \tensor{K}{_\mu_\nu^\s},
\label{Levi-Civita-gamma}
\end{align}
and it permits to write
\begin{align}
\nabla_{{\mu}} V^\nu &= \nabla^{(\g)}_{{\mu}} V^\nu + \tensor{K}{_{\mu}_\r^\nu} V^\r,
 \label{relation-contorsion-levi}
\end{align}
where $ \nabla^{(\g)}_{{\mu}} V^\nu = \partial_\mu  V^\nu + \tensor{\G}{^{(\g)}_\mu_\rho^\nu} V^\rho$. 

The Levi-Civita connection $\tensor{\G}{^{(G)}_\mu_\nu^\s}$  for the effective metric $G^{\mu\nu}$ is obtained as  
\begin{align}
  \tensor{\G}{^{(G)}_\mu_\nu^\s}
 &= \frac 12 G^{\s\r}\left(\del_\mu G_{\r\nu} 
   + \del_\nu G_{\r\mu}
  - \del_\r G_{\mu\nu}\right) 
  \nn\\
 &=\frac{1}{\rho}\left( \d^\s_\nu\del_\mu \r 
   +  \d^\s_\mu\del_\nu \r 
  -  \g_{\mu\nu}\g^{\s\r}\del_\r \r \right) 
  +  \frac 12 \g^{\s\r}\left(\del_\mu \g_{\r\nu} 
   + \del_\nu \g_{\r\mu}
  - \del_\r \g_{\mu\nu}\right) \ , 
\end{align}
which together with Eq. \eqref{Levi-Civita-gamma} gives
\begin{align}
\tensor{\G}{^{(G)}_\mu_\nu^\s}
 = \tensor{\widetilde\G}{_\mu_\nu^\s}  - \tensor{\cK}{_\mu_\nu^\s} \ .
\label{LC-contorsion-eff}
\end{align}
Here 
\begin{align}
 \tensor{\widetilde\G}{_\mu_\nu^\s} 
  &:= \tensor{\G}{_\mu_\nu^\s} +\frac{1}{\rho} \d^\s_\nu  \del_\mu \r \ , 
 \label{tilde-Gamma}
 \\
 \tensor{\cT}{_{\mu}_{\nu}^\s} &= \tensor{\widetilde \G}{_{\mu}_{\nu}^\s} - \tensor{\widetilde \G}{_{\nu}_{\mu}^\s} \ 
  = \tensor{T}{_{\mu}_{\nu}^\s} + \frac{1}{\rho} \left(\d_{\nu}^\s\del_{\mu}\rho - \d_{\mu}^\s\del_{\nu}\rho \right) \ ,
    \label{tilde-T-T}
\\
 \tensor{\cK}{_\mu_\nu^\s} &=\frac{1}{2}\left(\tensor{ \cT}{_{\mu}_{\nu}^{\s}}+\tensor{\cT}{^\sigma_\mu_\nu} -\tensor{\cT}{_\nu^\sigma_\mu}\right)=
   \tensor{K}{_{\mu}_{\nu}^{\s}}
  + \frac{1}{\rho}\left(G_{\mu\nu} G^{\sigma \rho}\del_\rho \r - \d^\s_{\mu}\del_\nu \r\right) 
   = -  \tensor{\cK}{_\mu^\s_\nu}
  \label{Levi-contorsion-full}
\end{align}
are the Weitzenb\"ock connection, the torsion, and the contorsion tensors of the effective frame, respectively \cite{Steinacker(2020)}. Hereafter, calligraphic fonts or a tilde indicate quantities related to the effective frame $\tensor{\mathcal{E}}{_a^\mu}$. 
The Weitzenb\"ock connection associated with the effective frame 
\begin{align} \label{nabla-tilde-math-E}
\widetilde\nabla_\nu \tensor{\cE}{_{a}^\mu} = 0
\end{align}
is  given explicitly using Eq. \eq{tilde-Gamma} by
\begin{align}
\widetilde\nabla_\mu V^{\s} &= \nabla_\mu V^{\s}
+ \left( \frac{1}{\rho}\del_{\mu}\r\right) V^\s 
= \nabla^{(G)}_\mu V^{\s} + \tensor{\cK}{_\mu_\k^\s} V^{\k}  \ , 
\nn\\
\widetilde\nabla_\mu V_\s  &= \nabla_\mu V_\s
- \left(\frac{1}{\rho}\del_{\mu}\r\right) V_\s
 = \nabla^{(G)}_\mu V_{\s} - \tensor{\cK}{_\mu_\s^\k} V_{\k}    \ ,
\label{nabla-nablatilde-rel}
\end{align}
where $ \nabla^{(G)}_{{\mu}} V^\nu = \partial_\mu  V^\nu + \tensor{\G}{^{(G)}_\mu_\rho^\nu} V^\rho$.

\section{Fermions in IKKT model}\label{Sec:fermions-in-IKKT}

In this section, we study the semi-classical geometric form of the Dirac-like action for fermions in the IKKT matrix model on a generic curved background. The discussion applies to generic noncommutative branes embedded through the first $3+1$ matrices as described in Sec. \ref{Sec:general-framework}\footnote{It turns out that the results also apply to $3+1$-dimensional branes with generic embedding in matrix models.}. This setup includes the case of covariant quantum spaces \cite{Sperling:2019xar,Steinacker(2020)}, which, in turn, encompass   the special FLRW   cosmic background which will be considered in Sec. \ref{Sec:Cosmic-solution}.

\subsection{Preliminaries}

We first establish the relation between the Cartan formulation of Riemannian geometry and the present framework based on the Weitzenb\"ock connection.
Let $\widehat{\omega}_{ab}=\widehat{\omega}_{\mu ab}\, dy^\mu =-\widehat{\omega}_{ba}$ be the torsion-free  Levi-Civita spin connection associated to the effective metric $G_{\mu \nu}$. Starting from the first Cartan structure equation \cite{Nakahara}
\begin{align}
    d \mathcal{E}^a = - \tensor{\widehat{\omega}}{^a_b} \wedge \mathcal{E}^b,
\end{align}
we obtain 
\begin{align}
\tensor{\mathcal{T}}{_\mu_\nu^a} = \tensor{\widehat{\omega}}{_\nu^a_b} \tensor{\mathcal{E}}{^b_\mu}-\tensor{\widehat{\omega}}{_\mu^a_b} \tensor{\mathcal{E}}{^b_\nu}=  \tensor{\widehat{\omega}}{_\nu^a_\mu}- \tensor{\widehat{\omega}}{_\mu^a_\nu},
\label{Cartan-structure-1}
\end{align}
where we have used the fact that the torsion of the Weitzenb\"ock connection is given by the
exterior derivative of the vielbein, which   yields  
\begin{align}
  \mathcal{T}^a= \frac{1}{2}\tensor{\mathcal{T}}{_\mu_\nu^a}\,  dy^\mu \wedge dy^\nu= \frac{1}{2}\left(\partial_\mu \tensor{\mathcal{E}}{^a_\nu}-\partial_\nu \tensor{\mathcal{E}}{^a_\mu}\right) dy^\mu \wedge dy^\nu.
\end{align}
The above relation represents the torsion two-form of the Weitzenb\"ock connection of the effective frame. 
Performing a cyclic permutation of the indices in Eq. \eqref{Cartan-structure-1}, we obtain 
\begin{align}
    \mathcal{K}_{\mu ab}=\widehat{\omega}_{\mu ab},
    \label{contortion&Levi-Civita-spin-connection}
\end{align}
which provides  the relation between the Levi-Civita spin connection 
and the  contorsion tensor  of the Weitzenb\"ock connection of the effective frame. 
This is easily seen to be consistent with the standard expression for the Levi-Civita spin connection
\begin{align}
\tensor{\widehat{\omega}}{_\mu^a^b} = \tensor{\mathcal{E}}{^{a}^\nu}\,
 \nabla^{(G)}_\mu  \tensor{\mathcal{E}}{^{b}_\nu}.
\label{Levi-Civita-spin-connection}
\end{align}
Of course, Eq. \eq{contortion&Levi-Civita-spin-connection} holds only for the effective frame $\cE$ underlying the Weitzenb\"ock connection, and does not allow local Lorentz transformations; the extension to general frames will be discussed in the next section.

\subsection{The Lagrangian}
\label{sec:Lagrangian}

The semi-classical action for a spinor in  Yang-Mills matrix models  can be written 
in arbitrary  local coordinates $y^\mu$ as (cf. Eq. \eqref{MM-action-IKKT})
\begin{align}
S = \Tr  \obar\Psi  \gamma_a [T^a,\Psi]
 \,\, \sim\,\, \int d^4 y\, \rho_M(y)\, \obar\Psi i \gamma_a E^{a\mu}
\partial_\mu \Psi \ .
 \label{fermionic-action-geom}
\end{align}
Here $T^a$ is the background solution of the matrix model, and the symbol $\sim$ indicates the semi-classical limit, where commutators are replaced by Poisson brackets. Moreover,  $\Psi$ is a matrix-valued spinor of $SO(D)$ (ignoring possible nonabelian gauge
fields  to simplify the notation), $\obar\Psi = \Psi^\dagger \gamma^{\widehat{0}} $ ($\gamma^{\widehat{0}}$ being the flat 0-th Dirac matrix, see Appendix \ref{Appendix:Dirac-matrices-conventions} for the conventions regarding Dirac matrices used in this paper), and $\r_M d^4 y$ is the symplectic volume form. 

In the special case of the IKKT model with $D=9+1$, the gamma matrices are those of $SO(9,1)$. 
We can then realize the aforementioned $3+1$-dimensional  spacetime in terms of the first 3+1 components $T^a$, setting the remaining $T^{A} =0$ for $A=4,...,9$.
The matrix model then reduces to noncommutative $\cN=4$ SYM on a $3+1$-dimensional spacetime brane $\cM^{3,1}$.
The  transversal directions 
will accommodate fuzzy extra dimensions, which are important for introducing mass terms (see Appendix   \ref{Sec:mass-term} for further details), as well as an induced Einstein-Hilbert action for gravity \cite{Steinacker:2021yxt}.

We note that the action \eqref{fermionic-action-geom} is written in the case of
Minkowski signature, whereas the Euclidean version involves the obvious
replacement $\obar\Psi \to \Psi^\dagger$. 
The (semi-classical) Lagrangian  in Eq. \eqref{fermionic-action-geom} can also be written as
\begin{align}
\mathcal{L} &= \frac{i}{2} \rho_M \left[ \obar{\Psi} \gamma^a \tensor{E}{_a^\mu}\partial_\mu \Psi - (\partial_\mu \obar{\Psi})  \gamma^a \tensor{E}{_a^\mu} \Psi \right] + i \rho \rho_M  m \obar{\Psi} \Psi,
\label{symmetrized-Lagrangian}    
\end{align}
where we have also introduced a mass term following  the line of reasoning  of Appendix   \ref{Sec:mass-term}. 

The most striking feature of this fermionic action is  that the spin connection seems to be \qm{missing} in the matrix Dirac operator
\begin{align}
\gamma_a [T^a,\Psi]\, 
\,\, \sim\,\, i \gamma_a E^{a\mu}\partial_\mu \Psi=i\gamma^a \tensor{E}{_a^\mu}\partial_\mu \Psi.
\label{Diracop-matrix}
\end{align}
However, 
we can rewrite the Lagrangian \eqref{symmetrized-Lagrangian} in terms of the standard covariant derivative for spinors, which reads as (see e.g. Refs. \cite{Nakahara,Gasperini-DeSabbata,DiGrezia2017})
\begin{align}
\widehat{D}_\mu \Psi 
= \left(\partial_\mu - \frac i2 \tensor{\widehat{\omega}}{_\mu^b^c} \,\Sigma_{bc}\right)\Psi, 
\label{covar-spinor}
\end{align}
where $\tensor{\widehat{\omega}}{_\mu^b^c}$   is the torsion-free  Levi-Civita spin connection
associated to the effective metric $G_{\mu \nu}$
 (see Eq. \eq{Levi-Civita-spin-connection}), and
\begin{align}\label{Sigma-a-b}
\Sigma_{ab} = \frac i4 [\gamma_a,\gamma_b] \, 
\end{align}
is the spinor representation of the generators of the Lorentz group.  Bearing in mind Eqs. \eqref{eff-frame-and-rho}, \eqref{contortion&Levi-Civita-spin-connection}, and \eqref{covar-spinor},   we find
\begin{align}
   \gamma^a \tensor{E}{_a^\mu} \partial_\mu \Psi = \rho \left( \gamma^a \tensor{\mathcal{E}}{_a^\mu} \widehat{D}_\mu \Psi + \frac{i}{2}\tensor{\mathcal{K}}{_\mu^b^c} \gamma^a \tensor{\mathcal{E}}{_a^\mu} \Sigma_{bc} \Psi \right) \ .
\end{align}
Using this expression, we can
rewrite the  Lagrangian   \eqref{symmetrized-Lagrangian} in the form 
\begin{align}
\mathcal{L}&=\frac{\mathcal{E}}{\rho} \left[ \frac{i}{2} \left( \obar\Psi \gamma^\mu \widehat{D}_\mu \Psi - (\widehat{D}_\mu \obar{\Psi}) \gamma^\mu \Psi    \right)  + i m \obar{\Psi} \Psi  -  \frac{1}{4}  \tensor{\mathcal{K}}{_\mu^b^c} \, \obar{\Psi} \, \{\gamma^\mu,\Sigma_{bc}\} \, \Psi  \right],
\label{Dirac-Lagrangian-1}
\end{align}
where  we have defined 
\begin{align}
    \gamma^\mu := \tensor{\mathcal{E}}{_a^\mu} \gamma^a, 
    \label{gamma-mu-curly-E}
\end{align}
and 
\begin{align}
\mathcal{E} :=  \det \left(\tensor{\mathcal{E}}{^{a}_\mu}\right) = \sqrt{-G} = \r_M \r^2,
\label{det-G-curly-E}  
\end{align}
with $ G  := \det(G_{\mu \nu})$. 
In terms of the Lagrangian \eqref{Dirac-Lagrangian-1}, the action of the spinor field reads 
\begin{align}
    S &= \int d^4y\, \cL. 
\end{align}

It is worth noting that the Eq. \eqref{Dirac-Lagrangian-1} mirrors, up to the factor $1/\rho$, the Dirac Lagrangian in a Riemann-Cartan spacetime \cite{Gasperini-DeSabbata}. In fact, upon working out the anticommutator $\{\gamma^\mu,\Sigma_{bc}\}$, it can be written as 
\begin{align}
\mathcal{L}&=\frac{\mathcal{E}}{\rho} \left[ \frac{i}{2} \left( \obar\Psi \gamma^\mu \widehat{D}_\mu \Psi - (\widehat{D}_\mu \obar{\Psi}) \gamma^\mu \Psi    \right)  + i m \obar{\Psi} \Psi 
-  \frac{i}{4}  \mathcal{K}_{[\alpha \beta \gamma]} \, \obar{\Psi} \gamma^{\alpha} \gamma^{\beta} \gamma^{\gamma} \Psi  \right].
\label{Dirac-Lagrangian-2}
\end{align}
Moreover, the totally antisymmetric contorsion term can be written 
on-shell (i.e., for backgrounds \eqref{background} which satisfy the equations of motion of the matrix model)  in terms of a gravitational axion $\tilde\rho$ as \cite{Fredenhagen:2021bnw}
\begin{align}
\mathcal{K}_{[\alpha \beta \gamma]} \gamma^{\alpha} \gamma^{\beta} \gamma^{\gamma} 
&= -\frac 16 \gamma_{\mu} \gamma_{\nu} \gamma_{\k}\varepsilon^{\mu\nu\k\s}  
\r^{-2}\del_{\s} \tilde\rho \ .
\end{align}
At this stage, it is useful to admit general  (non-parallel) frames $\tensor{e}{^a_\mu}$ via
\begin{equation}
    \eta_{a b} \tensor{e}{^a_\mu} \tensor{e}{^b_\nu} = G_{\mu \nu}
    \label{general-frame}
\end{equation}
so that the spinor $\Psi$ is allowed to transform as usual under local Lorentz transformations (we note that this step is only possible in the effective semi-classical description of the matrix model under consideration here, and allows a more convenient description of the fermionic action, similar as in teleparallel gravity \cite{Aldrovandi:2013wha}).
Correspondingly, we can introduce the following spin connection
\begin{align}
\tensor{\widetilde{\omega}}{_\mu^a^b} = e^{a \nu} \widetilde{\nabla}_\mu \tensor{e}{^b_\nu} = \tensor{\widehat{\omega}}{_\mu^a^b}-\tensor{\mathcal{K}}{_\mu^a^b},
\label{omega-tilde}    
\end{align}
where $\tensor{\widehat{\omega}}{_\mu^a^b} = e^{a \nu} \nabla^{(G)}_\mu \tensor{e}{^b_\nu} $ is the Levi-Civita spin connection associated to the general frame $\tensor{e}{^a_\nu}$  and $\tensor{\mathcal{K}}{_\mu^a^b}$ the contorsion tensor of the  Weitzenb\"ock connection $\tensor{\widetilde\G}{_\mu_\nu^\lambda} $ (note that we are employing for the Levi-Civita spin connection the same symbol as in Eq. \eqref{Levi-Civita-spin-connection}; this should not cause confusion, as henceforth we will always refer to the newly introduced $\tensor{\widehat{\omega}}{_\mu^a^b}$).
The associated  spinor covariant derivative is
\begin{align}
\widetilde{D}_\mu \Psi = \left(\partial_\mu - \frac i2 \tensor{\widetilde{\omega}}{_\mu^a^b} \,\Sigma_{ab}\right)\Psi = \widehat{D}_\mu \Psi + \frac{i}{2} \tensor{\mathcal{K}}{_\mu^a^b}\Sigma_{ab} \Psi, \label{tilde-D-mu}
\end{align}
 where $\widehat{D}_\mu \Psi$ can be read off from Eq. \eqref{covar-spinor}. This is nothing but the extension of the Weitzenb\"ock connection  to arbitrary frames;
note that   the spin connection \eqref{omega-tilde} vanishes in the physical frame due to Eq. \eqref{nabla-tilde-math-E}, i.e., when we make the replacement
\begin{align}
    \tensor{e}{^a_\mu} \rightarrow \tensor{\mathcal{E}}{^a_\mu}.
\end{align}
By means of the  formulae \eqref{general-frame}--\eqref{tilde-D-mu}, the Lagrangian function \eqref{Dirac-Lagrangian-2} assumes, in the general frame $\tensor{e}{^a_\mu}$,  the form
\begin{align}
\mathcal{L}&=\frac{e}{\rho} \left[ \frac{i}{2} \left( \obar\Psi \gamma^\mu \widetilde{D}_\mu \Psi - (\widetilde{D}_\mu \obar{\Psi}) \gamma^\mu \Psi    \right)  + i m \obar{\Psi} \Psi\right]
\nonumber \\
&=\frac{e}{\rho} \left[ \frac{i}{2} \left( \obar\Psi \gamma^\mu \partial_\mu \Psi - (\partial_\mu \obar{\Psi}) \gamma^\mu \Psi    \right)  + i m \obar{\Psi} \Psi  +  \frac{i}{4}  \widetilde{\omega}_{[\alpha \beta \gamma]} \, \obar{\Psi} \gamma^{\alpha} \gamma^{\beta} \gamma^{\gamma} \Psi  \right],
\label{Dirac-Lagrangian-3}
\end{align}
 where, similarly to Eqs. \eqref{gamma-mu-curly-E} and \eqref{det-G-curly-E}, 
 \begin{subequations}
\begin{align}
\gamma^\mu &:= \tensor{e}{_a^\mu} \gamma^a,
\label{gamma-matrix-general-frame}
 \\
e &:=  \det \left(\tensor{e}{^{a}_\mu}\right)= \sqrt{-G}.
\end{align}
\end{subequations}
Note that the Lagrangian \eqref{Dirac-Lagrangian-3} reduces to Eq. 
\eqref{symmetrized-Lagrangian} for the parallel frame $\tensor{\mathcal{E}}{^a_\mu}$, where $\tensor{\widetilde{\omega}}{_\mu^a^b}$ vanishes. Furthermore, it is worth pointing out that we have used for the Dirac matrices the same notation as in Eq. \eqref{gamma-mu-curly-E}; no confusion should arise since from now on we will consider the matrices defined in Eq. \eqref{gamma-matrix-general-frame}. As a consequence of  Eq.  \eqref{Dirac-Lagrangian-3},  the equations of motion read as
\begin{align}
\gamma^\mu \widehat{D}_\mu \Psi + m \Psi -\frac{1}{4} \mathcal{K}_{[\alpha \beta \gamma]}\gamma^\alpha \gamma^\beta\gamma^\gamma \Psi + \frac{\rho}{2} \left(\partial_\mu \rho^{-1}\right) \gamma^\mu \Psi=0 \ ,  
\label{quantum-Dirac-equation}
\end{align}
where the last terms breaks the local Lorentz invariance
on non-trivial backgrounds.

It follows from  the Dirac equation \eqref{quantum-Dirac-equation} and the equality\footnote{It is worth pointing out that also the relation  $\widetilde{D}_\mu \gamma^\alpha =0$ holds. }
\begin{align}
\widehat{D}_\mu \gamma^\alpha =0,    
\end{align}
that the effective current $\r^{-1} J^\m$ is conserved, i.e., 
\begin{align}
 \nabla^{(G)}_\mu \left(\r^{-1} J^\mu\right)=0, 
 \label{conservation-eff-J-mu}
\end{align}
where $J^\mu := \obar{\Psi} \gamma^\mu \Psi$.

\section{The particle limit of the Dirac field}
\label{Sec:particle-limit-Dirac}

In this section, we will work out the particle limit of the Dirac field by applying the JWKB approximation to the quantum-mechanical  Dirac equation \eqref{quantum-Dirac-equation}. This means that we assume the validity of the semi-classical limit, where the particle is characterized by a world line and its spin by a polarization vector, and the gravitational field is supposed to be slowly varying. 

The  classical motion of the fermionic field is dealt with in  Sec. \ref{Sec:classical-trajectory}, while Sec. \ref{Sec:quantum-motion} is devoted to the study of the quantum dynamics. 

\subsection{The classical trajectory}\label{Sec:classical-trajectory}

Following the recipe of the JWKB scheme (see e.g. Refs. \cite{Rudiger1981,Audretsch1981,Audretsch1981b,Hayashi1990,Nomura1992,Cianfrani2008,Khanapurkar2018}),  we adopt the \emph{ansatz} where the  solution $\Psi$ of the Dirac equation can be written as a phase factor and a spinor amplitude via the following series:
\begin{align}
\Psi(x)= \exp \left(- \frac{i}{\hbar} W(x)\right) \sum_{n=0}^{\infty} \hbar^n \psi^{(n)}(x),
\label{JWKB-approx}
\end{align}
where  $W(x)$ is a real-valued function and $\psi^{(n)}(x)$ a spinor. If we insert the above formula in Eq. \eqref{quantum-Dirac-equation} and equate the coefficients involving the same powers of $\hbar$, we obtain at leading order and to next-to-leading order
\begin{subequations}
\begin{align}
&\left(i \gamma^\mu \partial_\mu W -m\right)\psi^{(0)}=0,
\label{Order-zero-JWKB}
\\
&\left(i \gamma^\mu \partial_\mu W -m\right)\psi^{(1)}=\gamma^\mu \widehat{D}_{\mu} \psi^{(0)}-\frac{1}{4}\mathcal{K}_{[\alpha \beta \gamma]} \gamma^\alpha \gamma^\beta \gamma^\gamma \psi^{(0)} +\frac{\rho}{2} \left(\partial_\mu \rho^{-1}\right)\gamma^\mu \psi^{(0)},
\label{Order-one-JWKB}
\end{align}
\end{subequations}
respectively. Note that in order to obtain  the above equations it is necessary to replace  the mass $m$ in Eq. \eqref{quantum-Dirac-equation}  by $m/\hbar$.

The solvability condition $\det (i \gamma^\mu \partial_\mu W-m)=0$ of Eq. \eqref{Order-zero-JWKB} implies the Hamilton-Jacobi equation for a relativistic nonspinning particle
\begin{align}
    G^{\mu \nu}p_\mu p_\nu =-m^2,
\end{align}
where $p_\mu = - \partial_\mu W$. The normalized timelike vector
\begin{subequations}
\begin{align}
& u_\alpha = \frac{-\partial_\alpha W}{\vert G^{\mu \nu} \partial_\mu W \partial_\nu W \vert^{1/2}}=\frac{1}{m} p_\alpha,
\label{u-alpha-p-alpha}
\\
& G^{\mu \nu} u_\mu u_\nu =-1,
\label{u-mu-u-mu}
\end{align}
\end{subequations}
represents the tangent vector (i.e., the four-velocity) to the worldlines orthogonal to the family of spacelike hypersurfaces  $W=constant$ having constant phase. By standard arguments \cite{Carroll2004}, one can prove that these trajectories form a congruence of timelike geodesics
\begin{equation}
u^\alpha \nabla^{(G)}_\alpha u^\beta=0,
\label{geodesic-congruence}    
\end{equation}
which is rotation free
\begin{align}
    \Omega_{\alpha \beta} := \nabla^{(G)}_{[\beta} u_{\alpha]}=0.
\end{align}
Therefore, to zero order in $\hbar$, we obtain the completely classical result according to which the motion of the Dirac fermion is not influenced by the spin, i.e., the particle follows a geodesic trajectory  of the background geometry. The remaining kinematical properties of the geodesic congruence are embodied by
\begin{align}
\label{geodesic-congruence-kinematics}
    \nabla^{(G)}_{\beta} u_{\alpha}= \frac{1}{3} \widehat{\theta} P_{\alpha \beta} + \widehat{\sigma}_{\alpha \beta},
\end{align}
where
\begin{subequations}
\begin{align}
 \widehat{\sigma}_{\alpha \beta} &=    \nabla^{(G)}_{(\beta} u_{\alpha)} - \frac{1}{3} \widehat{\theta} P_{\alpha \beta}, 
 \\
 \widehat{\theta} &=  \nabla^{(G)}_{\beta} u^{\beta}, 
 \\
 P_{\alpha \beta} &= G_{\alpha \beta} + u_\alpha u_\beta,
\end{align}
\end{subequations}
represent the shear tensor, the expansion scalar, and the transverse metric (fulfilling the role of a projector onto the space orthogonal to $u^\alpha$), respectively.  

It follows from Eq. \eqref{Order-zero-JWKB} that the spinor $\psi^{(0)}$ describes the positive-energy solutions of the flat-space Dirac equation and hence it assumes the general form
\begin{align}
\psi^{(0)}(x) = \beta_1(x) u^{(1)}(x) + \beta_2(x) u^{(2)}(x),    \qquad \beta_1(x),\beta_2(x) \in \mathbb{C}, 
\label{psi-0-u1-u2}
\end{align}
where the spin-up and spin-down spinors are, in the Dirac basis\footnote{Here the $SO(9,1)$ spinors  of the matrix model  are decomposed in terms of 3+1-dimensional spinors, as explained in Appendix \ref{Sec:mass-term}.}, \cite{Peskin1995}
\begin{subequations}
\label{spinors-u1-u2-general}
\begin{align}
    u^{(1)} &= \left(\frac{p^{\widehat{0}}+m}{2m}\right)^{1/2} \begin{bmatrix}
  1 \\ 0 \\ p^{\widehat{3}}/(p^{\widehat{0}} +m) \\(p^{\widehat{1}}+i p^{\widehat{2}})/(p^{\widehat{0}} +m)
 \end{bmatrix},
\\
 u^{(2)} &= \left(\frac{p^{\widehat{0}}+m}{2m}\right)^{1/2} \begin{bmatrix}
  0 \\ 1 \\ (p^{\widehat{1}}-i p^{\widehat{2}})/(p^{\widehat{0}} +m) \\ -p^{\widehat{3}}/(p^{\widehat{0}} +m) 
 \end{bmatrix},
\end{align}
\end{subequations}
respectively, and $p^a = \tensor{e}{^a_\mu} p^\mu$. 

The condition for the existence of a nontrivial solution $\psi^{(1)}$ of Eq. \eqref{Order-one-JWKB} is that all solutions of the corresponding transposed homogeneous equation are orthogonal to the inhomogeneity (Fredholm  alternative, see Refs. \cite{Audretsch1981,Audretsch1981b,Alsing2009} for further  details). Therefore, the solvability conditions of Eq. \eqref{Order-one-JWKB} yield 
\begin{subequations}
\label{solvability-1}
\begin{align}
& \obar{u}^{(1)}  \left[ \gamma^\mu \widehat{D}_{\mu} \psi^{(0)}-\frac{1}{4}\mathcal{K}_{[\alpha \beta \gamma]} \gamma^\alpha \gamma^\beta \gamma^\gamma \psi^{(0)} + \frac{\rho}{2} \left(\partial_\mu \rho^{-1}\right) \gamma^\mu \psi^{(0)}\right]=0,
\\
& \obar{u}^{(2)}  \left[ \gamma^\mu \widehat{D}_{\mu} \psi^{(0)}-\frac{1}{4}\mathcal{K}_{[\alpha \beta \gamma]} \gamma^\alpha \gamma^\beta \gamma^\gamma \psi^{(0)} + \frac{\rho}{2} \left(\partial_\mu \rho^{-1}\right) \gamma^\mu \psi^{(0)}\right]=0,
\end{align}
\end{subequations}
where we have exploited Eq. \eqref{psi-0-u1-u2}. 

At this stage, we restrict our attention to an arbitrary but fixed worldline of the geodesic congruence admitting $u^\alpha$ as the tangent vector field (cf. Eq. \eqref{geodesic-congruence}). In this setting, we can write on the worldline
\begin{subequations}
\label{rest-frame-1,2,3}
\begin{align}
e_{\widehat{0}}^{\ \alpha}   &\overset{\star}{=} u^\alpha, 
\label{rest-frame-1} 
\\
u^\mu \nabla^{(G)}_\mu \tensor{e}{_a^\alpha} &\overset{\star}{=}0,
\label{rest-frame-2} 
\\
\tensor{\widehat{\omega}}{_\mu^a^b} &\overset{\star}{=} 0, 
\label{rest-frame-3}
\end{align}
\end{subequations}
the star  symbol standing for an equality valid on the worldline (we will omit the star if an equation is valid in any frame). In Eq. \eqref{rest-frame-1}, we have adjusted the vector $e_{\widehat{0}}^{\ \alpha}$ parallel to the velocity $u^\alpha$; in Eq. \eqref{rest-frame-2}, we have parallelly propagated the tetrad  along the chosen $u^\alpha$ direction so that, consistently with Eq. \eqref{geodesic-congruence}, the covariant derivative (with respect to the  the Christoffel symbols $\tensor{\G}{^{(G)}_\mu_\nu^\lambda}$) of $\tensor{e}{_a^\alpha}$ vanishes on the worldline; last, Eq. \eqref{rest-frame-3} stems from Eq. \eqref{rest-frame-2}. It is thus clear that  the choice \eqref{rest-frame-1,2,3}  amounts to introducing the particle's rest frame and the related Fermi normal coordinates \cite{MTW,Hartle2003gravity}.

In the particle's rest frame, the four-momentum $p^a$ is such that  $p^a \overset{\star}{=}(m,\boldsymbol{0})$ and hence Eq. \eqref{spinors-u1-u2-general} reduces to the rest-frame positive-energy Dirac spinors
\begin{subequations}
\label{spinors-u1-u2-rest-frame}
\begin{align}
    u^{(1)} &\overset{\star}{=} \begin{bmatrix}
  1 \\ 0 \\0 \\ 0
 \end{bmatrix},
\\
 u^{(2)} &\overset{\star}{=}\begin{bmatrix}
  0 \\ 1 \\ 0 \\ 0
 \end{bmatrix}.
\end{align}
\end{subequations}
Furthermore, bearing in mind Eqs. \eqref{geodesic-congruence} and \eqref{geodesic-congruence-kinematics} jointly with Eqs. \eqref{rest-frame-1,2,3} and \eqref{spinors-u1-u2-rest-frame}, we obtain the following relations:
\begin{subequations}
\begin{align}
\tensor{e}{_b^\alpha} \partial_\alpha p^{\widehat{0}} &\overset{\star}{=} 0,
\label{relation-e-del-p-1}
 \\
e_{\widehat{0}}^{\ \alpha} \partial_\alpha p^{\widehat{a}} &\overset{\star}{=} 0,
\label{relation-e-del-p-2}
 \\
e_{\widehat{b}}^{\ \alpha} \partial_\alpha p^{\widehat{a}} &\overset{\star}{=} m \ e^{\widehat{a} \epsilon} \ e_{\widehat{b}}^{\ \alpha} \left( \widehat{\sigma}_{\epsilon \alpha} + \frac{1}{3} \widehat{\theta} P_{\epsilon \alpha}\right).
\end{align}
\end{subequations}
Moreover, from Eq. \eqref{spinors-u1-u2-rest-frame},  (herafter, $A,B=1,2$)
\begin{subequations}
\label{relation-u-gamma-u}
\begin{align}
\obar{u}^{(A)} \gamma^\mu u^{(A)}  &\overset{\star}{=} -u^\mu,
 \\
\obar{u}^{(A)} \gamma^\mu u^{(B)}  &\overset{\star}{=} 0, \qquad (A \neq B),
\end{align}
\end{subequations}
and 
\begin{subequations}
\label{relation-u-del-gamma-u}
\begin{align}
\obar{u}^{(A)} \gamma^\mu \partial_\mu u^{(A)}  &\overset{\star}{=} -\widehat{\theta}/2,
\\
\obar{u}^{(A)} \gamma^\mu \partial_\mu u^{(B)}   &\overset{\star}{=} 0, \qquad (A \neq B).
\end{align}
\end{subequations}
Last, owing to Eqs. \eqref{relation-e-del-p-1} and  \eqref{relation-e-del-p-2},
\begin{align}
u^\alpha \widetilde{D}_\alpha u^{(A)}&\overset{\star}{=}-\frac{1}{4} u^\alpha \mathcal{K}_{\alpha \beta \gamma} \gamma^\beta \gamma^\gamma  u^{(A)},
\label{u-tilde-D-u-A}
\end{align}
which leads to the generally valid relation
\begin{align}
     u^\alpha \widehat{D}_\alpha u^{(A)} =0,
\end{align}
upon taking into account Eq. \eqref{rest-frame-3}. 

\subsection{The quantum dynamics}\label{Sec:quantum-motion}

At this stage, we have all the ingredients to evaluate the quantum corrections to the fermionic dynamics, i.e., the corrections due to the wave-like nature of the the fermions. After some preliminary calculations, the spin precession equation  and the translation motion will be worked out in Secs. \ref{Sec:spin-precession-equation} and \ref{Sec:translational-motion}, respectively. Last, we evaluate the magnetic dipole moment of the Dirac particle in Sec. \ref{Sec:magnetic-dipole}.

Upon using Eqs. \eqref{psi-0-u1-u2}, \eqref{rest-frame-3}, \eqref{relation-u-gamma-u}, and \eqref{relation-u-del-gamma-u}, the solvability condition \eqref{solvability-1} leads to generally valid equations
\begin{subequations}
\label{u-mu-del-beta-1,2}
\begin{align}
u^\mu \partial_\mu \beta_1  = -\frac{\widehat{\theta}}{2} \beta_1 -\frac{1}{4} \mathcal{K}_{[\alpha \beta \gamma]} \biggl( \obar{u}^{(1)} \gamma^\alpha \gamma^\beta \gamma^\gamma \beta_1 u^{(1)} +\obar{u}^{(1)} \gamma^\alpha \gamma^\beta \gamma^\gamma \beta_2 u^{(2)}  \biggr)-\frac{\rho}{2} \left(\partial_\mu \rho^{-1}\right) \beta_1 u^\mu,
\\
u^\mu \partial_\mu \beta_2  = -\frac{\widehat{\theta}}{2} \beta_2 -\frac{1}{4} \mathcal{K}_{[\alpha \beta \gamma]} \biggl( \obar{u}^{(2)} \gamma^\alpha \gamma^\beta \gamma^\gamma \beta_1 u^{(1)} +\obar{u}^{(2)} \gamma^\alpha \gamma^\beta \gamma^\gamma \beta_2 u^{(2)}  \biggr)-\frac{\rho}{2} \left(\partial_\mu \rho^{-1}\right) \beta_2 u^\mu,
\end{align}
\end{subequations}
describing the propagation of the scalar functions $\beta_1,\beta_2$ along the geodesic trajectory.  Therefore,  the propagation equation for the spinor $\psi^{(0)}$ can be obtained starting from Eqs.  \eqref{u-tilde-D-u-A} and \eqref{u-mu-del-beta-1,2}, and reads as
\begin{align} \label{prop-eq-psi-0}
    u^\alpha \widetilde{D}_\alpha \psi^{(0)} = -\frac{\widehat{\theta}}{2}\psi^{(0)} - \frac{i}{2} \mathcal{K}_{[\alpha \beta] \gamma} \sigma^{\alpha \beta} u^\gamma \psi^{(0)} -\frac{\rho}{2} \left(\partial_\mu \rho^{-1}\right)  u^\mu  \psi^{(0)},
\end{align}
where (cf. Eq. \eqref{Sigma-a-b})
\begin{align}
    \sigma^{\alpha \beta} = 2 \Sigma^{\alpha \beta}.
\end{align}
In deriving Eq. \eqref{prop-eq-psi-0}, we have also taken into account that
\begin{align}
u^{(1)} \obar{u}^{(1)} + u^{(2)} \obar{u}^{(2)} =-\frac{\gamma^\mu p_\mu + i m}{2m} \equiv A_{+},
\end{align}
with
\begin{align}
i   A_{+} A_{+} &= A_{+},
 \nn \\
A_{+} u^{(A)}&=- i u^{(A)}. 
\end{align}

At this stage, let us introduce the normalized spinor $b^{(0)}(x)$ via the relations
\begin{align}
\psi^{(0)}(x)=f(x) b^{(0)}(x),
 \nn \\
i \obar{b}^{(0)} b^{(0)}=1,
\label{psi-0-f-b-0}
\end{align}
where the real-valued function $f(x)$ satisfies (cf. Eq. \eqref{psi-0-u1-u2})
\begin{align}
f^2(x)= \vert \beta_1(x) \vert^2 + \vert \beta_2 (x)\vert^2.
\label{f-squared}    
\end{align}
Then, if we  employ Eqs. \eqref{u-mu-del-beta-1,2} and \eqref{f-squared}, we find for the function $f(x)$  the propagation equation
\begin{align} \label{prop-equat-f}
    u^\mu \partial_\mu f = -\frac{\widehat{\theta}}{2} f -\frac{\rho}{2} \left(\partial_\mu \rho^{-1}\right) u^\mu f, 
\end{align}
whereas  for the normalized spinors $b^{(0)}$ and $\obar{b}^{(0)}$ we can write
\begin{subequations}
\begin{align}
u^\alpha \widetilde{D}_\alpha  b^{(0)} &= -\frac{i}{2} \mathcal{K}_{[\alpha \beta]\gamma} \sigma^{\alpha \beta} u^\gamma b^{(0)},
\\
u^\alpha \widetilde{D}_\alpha  \obar{b}^{(0)} &= \frac{i}{2} \mathcal{K}_{[\alpha \beta]\gamma} \obar{b}^{(0)} \sigma^{\alpha \beta} u^\gamma,
\end{align}
\end{subequations}
once  Eqs.  \eqref{prop-eq-psi-0} and \eqref{prop-equat-f} have been exploited. 

\subsubsection{The spin precession equation}\label{Sec:spin-precession-equation}

It will prove to be useful the introduction of a new connection. Following Ref. \cite{Audretsch1981b}, we define the new affinities  $\tensor{\overstar{\Gamma}}{_\mu_\nu^\lambda}$ and $\tensor{\overstar{\omega}}{_\mu^a^b}$ as 
\begin{align}
\tensor{\overstar{\Gamma}}{_\mu_\nu^\lambda} &= \tensor{\widetilde{\Gamma}}{_\mu_\nu^\lambda} + 2 \mathcal{K}_{[\nu \epsilon]\mu}G^{\epsilon \lambda} =  \tensor{\G}{^{(G)}_\mu_\nu^\lambda} + 3 \mathcal{K}_{[\mu \nu \epsilon]}G^{\epsilon \lambda}, 
\label{Gamma-star}
 \\
\tensor{\overstar{\omega}}{_\mu^a^b} &= \tensor{\widetilde{\omega}}{_\mu^a^b}  -2 \mathcal{K}^{[ab]}_{\ \ \ \ \mu} = \tensor{\widehat{\omega}}{_\mu^a^b} -3 \mathcal{K}^{[ab \epsilon]}G_{\epsilon \mu},
\label{omega-star}
\end{align}
with
\begin{align}
 \tensor{\overstar{\Gamma}}{_\mu_\nu^\lambda} &= \tensor{e}{_a^\lambda}    \overstar{D}_\mu \tensor{e}{^a_\nu} = \tensor{e}{_a^\lambda}   \left( \partial_\mu \tensor{e}{^a_\nu} + \tensor{\overstar{\omega}}{_\mu^a_b} \tensor{e}{^b_\nu}\right),
 \\
 \tensor{\overstar{\omega}}{_\mu^a^b} &= e^{a \nu} \overstar{\nabla}_\mu \tensor{e}{^b_\nu} = e^{a \nu}\left(\partial_\mu\tensor{e}{^b_\nu} - \tensor{\overstar{\Gamma}}{_\mu_\nu^\lambda} \tensor{e}{^b_\lambda}\right).
\end{align}
The new connection is compatible with the effective  metric, as $\overstar{\nabla}_\alpha G_{\mu \nu}=0$, and satisfies the following relations
\begin{subequations}
\begin{align}
V^\epsilon    \overstar{\nabla}_\epsilon V^\alpha &= V^\epsilon \nabla^{(G)}_\epsilon V^\alpha,
\label{relation-nabla-star-1}
\\
\overstar{\nabla}_\alpha V^\alpha &=  \nabla^{(G)}_\alpha V^\alpha, 
\label{relation-nabla-star-2}
\\
\overstar{D}_\mu \gamma^\alpha &=0,
\label{nabla-star-gamma}
\end{align}
\end{subequations}
 $V^\alpha$ being a generic vector. 

Bearing in mind Eqs. \eqref{prop-eq-psi-0} and \eqref{omega-star}, we find for the spinor $\psi^{(0)}$
\begin{align}
    u^\alpha \overstar{D}_\alpha \psi^{(0)}= - \frac{\widehat{\theta}}{2} \psi^{(0)} -\frac{\rho}{2} \left(\partial_\mu \rho^{-1}\right) u^\mu \psi^{(0)},
\end{align}
which, in turn, implies that
\begin{align}
u^\alpha \overstar{D}_\alpha b^{(0)} &=0 \ ,
 \nn \\
u^\alpha \overstar{D}_\alpha \obar{b}^{(0)}&=0 \ ,
\label{D-star-b-b-bar}
\end{align}
upon exploiting the propagation equation \eqref{prop-equat-f}. In other words, the normalized spinors $ b^{(0)}$ and $\obar{b}^{(0)}$ are parallelly propagated along the geodesic path, which  represents the trajectory followed by the particle in the completely classical limit (see Eq. \eqref{Order-zero-JWKB}), provided that we employ the new connections  $\tensor{\overstar{\Gamma}}{_\mu_\nu^\lambda}$ and $\tensor{\overstar{\omega}}{_\mu^a^b}$. 

The spin vector of the Dirac particle can be written via the JWKB approximation as (see e.g. Refs. \cite{Audretsch1981,Audretsch1981b} for further details) 
\begin{align}
    S^\alpha= S^\alpha_{(0)} + {\rm O}\left(\hbar\right),
\end{align}
the lowest-order correction being
\begin{align}
S^\alpha_{(0)} = \frac{1}{2} \varepsilon^{\alpha \beta \gamma \delta} u_{\beta} \obar{b}^{(0)} \sigma_{\gamma \delta} b^{(0)},
\label{spin-vector-1}
\end{align}
with
\begin{align}
\varepsilon^{\alpha \beta \gamma \delta} = \tensor{e}{_a^\alpha}    \tensor{e}{_b^\beta}    \tensor{e}{_c^\gamma}    \tensor{e}{_d^\delta} \epsilon^{abcd},    
\end{align}
where the totally antisymmetric Levi-Civita symbol $\epsilon^{abcd}$ is such that, 
in our conventions, $\epsilon^{0123}=1$. The spin vector \eqref{spin-vector-1} satisfies
\begin{align}
u_\alpha   S^\alpha_{(0)}&=0, 
\nn \\
G_{\alpha \beta} S^\alpha_{(0)}S^\beta_{(0)}&=1,
 \end{align}
and is characterized by the propagation equation 
\begin{align}
u^\mu \overstar{\nabla}_\mu S^{\alpha}_{(0)}   =0,
\label{nabla-star-spin}
\end{align}
which can be established by means of Eqs.  \eqref{geodesic-congruence}, \eqref{relation-nabla-star-1},   \eqref{nabla-star-gamma}, and  \eqref{D-star-b-b-bar},    jointly with the identity $\overstar{\nabla}_\mu \varepsilon^{\alpha \beta \gamma \delta}=0$. Therefore, through the new connections \eqref{Gamma-star} and \eqref{omega-star}, the lowest-order spin vector \eqref{spin-vector-1} is parallely transported along the particle's classical geodesic trajectory. In terms of the Levi-Civita connection (see Eq. \eqref{Gamma-star}) and the axial-vector part of the contorsion tensor
\begin{align}
\mathcal{A}^\mu = \frac{1}{6} \varepsilon^{\mu \alpha \beta \gamma}  \mathcal{K}_{[\alpha \beta \gamma]},
\end{align}
Eq. \eqref{nabla-star-spin} implies the spin precession equation 
\begin{align}
u^\rho \nabla^{(G)}_\rho  S^{\mu}_{(0)} = 3 \varepsilon^{\mu \alpha \lambda \epsilon} \mathcal{A}_\alpha S_{(0)\lambda} u_\epsilon.
\label{spin-precession}
\end{align}

\subsubsection{The nongeodesic translational motion}\label{Sec:translational-motion}

Let us introduce the Gordon decomposition of the effective Dirac current 
\begin{align}
  \rho^{-1}  J^\mu \equiv \mathcal{J}^{\mu}_{\rm M}+ \mathcal{J}^{\mu}_{\rm C} \ ,
\end{align}
where the magnetization and convection currents can be obtained starting from the  Dirac equation \eqref{quantum-Dirac-equation} and the identity \eqref{omega-star}. Explicitly,  $\mathcal{J}^{\mu}_{\rm M} $  and $\mathcal{J}^{\mu}_{\rm C}$ read as, respectively,
\begin{subequations}
\begin{align}
\mathcal{J}^{\mu}_{\rm M} &= \frac{i \hbar}{2m\rho}   \left[\widehat{D}_{\nu} \left(\obar{\Psi} \sigma^{\mu \nu} \Psi\right) + \rho \left(\partial_\nu \rho^{-1}\right) \obar{\Psi} \sigma^{\mu \nu} \Psi\right] =\frac{i \hbar}{2m}  \widehat{D}_{\nu} \left(\frac{\obar{\Psi} \sigma^{\mu \nu} \Psi}{\rho}\right),  
\label{magnetization-current}
\\
 \mathcal{J}^{\mu}_{\rm C} &= \frac{ \hbar}{2m\rho} \left[\left(\widehat{D}^\mu \obar{\Psi}\right)\Psi -\obar{\Psi} \widehat{D}^\mu \Psi - \frac{3i}{2} \mathcal{K}_{[\alpha \beta \gamma]}\obar{\Psi} \sigma^{\alpha \beta} G^{\gamma \mu} \Psi\right] = \frac{ \hbar}{2m\rho} \left[\left(\overstar{D}^\mu \obar{\Psi}\right)\Psi -\obar{\Psi} \overstar{D}^\mu \Psi\right].
\end{align}
\end{subequations}
By means of the the commutation relations for the covariant derivative operator $\widehat{D}_\mu$ and Eq. \eqref{conservation-eff-J-mu}, it can be shown that these quantities are conserved, i.e., they satisfy
\begin{subequations}
\begin{align}
\nabla^{(G)}_\mu   \mathcal{J}^{\mu}_{\rm M} &=0,
\label{D-mu-J-M}
\\
\nabla^{(G)}_\mu   \mathcal{J}^{\mu}_{\rm C} &=0.
\end{align}
\end{subequations}
In particular, the torsion-free condition featuring the operator $\widehat{D}_\mu$ is essential for the  evaluation of Eq. \eqref{D-mu-J-M}, as it  guarantees that  $[\widehat{D}_\mu,\widehat{D}_\nu]\rho^{-1}=0$. Physically, $\mathcal{J}^{\mu}_{\rm M}$ represents the curl of the spin density and can be interpreted as a magnetization current, while $\mathcal{J}^\mu_{\rm C}$ is a convection four-current as  its spacelike part resembles the three-vector probability current of Schr\"{o}dinger theory \cite{Audretsch1981,Audretsch1981b}. 

Due to its physical interpretation, the convection current $\mathcal{J}^\mu_{\rm C}$ can be used  to define the particle's translational motion. Thus, we can define a congruence of timelike curves having tangent vector $v^\alpha$, which is  given by
\begin{align}
    v^\alpha= \frac{\mathcal{J}^\alpha_{\rm C}}{\sqrt{-G_{\mu \nu}\mathcal{J}^{\mu}_{\rm C}\mathcal{J}^{\nu}_{\rm C}}},
\end{align}
which upon exploiting Eqs. \eqref{JWKB-approx}, \eqref{u-mu-u-mu}, \eqref{psi-0-f-b-0}, and  \eqref{D-star-b-b-bar}, yields
\begin{align}
    v^\mu = u^\mu + \frac{\hbar}{2m} \left[ \left(\overstar{D}^\mu \obar{b}^{(0)}\right)b^{(0)}-\obar{b}^{(0)} \overstar{D}^{\mu} b^{(0)}\right]+ {\rm O}\left(\hbar^2 \right).
\end{align}
The above formula shows that the spin forces the particle to follow a quantum corrected trajectory which  deviates from the geodesic motion, which is pursued only at classical level (see Eq. \eqref{geodesic-congruence}). In fact, we can evaluate the nongeodesic acceleration $a_\alpha$ of the fermion as follows. Let us start with the following expression: 
\begin{align}
a_\alpha &= v^\beta \nabla^{(G)}_\beta v_\alpha= v^\beta \overstar{\nabla}_\beta v_\alpha=2v^\beta \overstar{\nabla}_{[\beta} v_{\alpha]}
\nonumber \\
&=\frac{\hbar}{m} u^\beta \left[\left(\overstar{D}_{[\beta}\overstar{D}_{\alpha]} \obar{b}^{(0)}\right)b^{(0)}-\obar{b}^{(0)}\left(\overstar{D}_{[\beta}\overstar{D}_{\alpha]} b^{(0)}\right)\right]-2v^\beta \overstar{\Gamma}_{[\beta \alpha]}^{\;\;\; \;\;\;\, \lambda} u_\lambda +{\rm O}\left(\hbar^2 \right),
\end{align}
where we have exploited Eqs. \eqref{u-alpha-p-alpha},  \eqref{relation-nabla-star-1}, and \eqref{D-star-b-b-bar} jointly with the normalization condition $v^\alpha v_\alpha=-1$. The above formula can be further simplified by exploiting Eq. \eqref{Gamma-star} and the commutation relations 
\begin{align}
\left[\overstar{D}_\mu,\overstar{D}_\nu\right]\Psi &= -\frac{i}{4} \overstar{R}_{\mu \nu}^{\;\;\;\; ab} \sigma_{ab} \Psi - 2 \overstar{\Gamma}_{[\mu \nu]}^{\;\;\;\;\;\; \lambda} \overstar{D}_{\lambda} \Psi,
\nn \\
\left[\overstar{D}_\mu,\overstar{D}_\nu\right]\obar{\Psi} &= \frac{i}{4} \overstar{R}_{\mu \nu}^{\;\;\;\; ab} \obar{\Psi} \sigma_{ab}  - 2 \overstar{\Gamma}_{[\mu \nu]}^{\;\;\;\;\;\; \lambda} \overstar{D}_{\lambda} \obar{\Psi},
\end{align}
where
\begin{align}
\overstar{R}_{\mu \nu}^{\;\;\;\; ab} = \partial_\mu \overstar{\omega}_\nu^{\;\; ab}    - \partial_\nu \overstar{\omega}_\mu^{\;\; ab}   +  \overstar{\omega}_\mu^{\;\; ac} \overstar{\omega}_{\nu c}^{\;\;\;\, b}-\overstar{\omega}_\nu^{\;\; ac} \overstar{\omega}_{\mu c}^{\;\;\;\, b}.
\end{align}
In this way, we end up with the final form of the acceleration vector describing a nongeodesic motion to first order in $\hbar$
\begin{align}
a_\alpha= -\frac{i}{2}\left(\frac{ \hbar}{2m}\right) \overstar{R}_{\alpha \beta \mu \nu} u^\beta \, \obar{b}^{(0)} \sigma^{\mu \nu} b^{(0)} +{\rm O}\left(\hbar^2 \right),
\label{acceleration}
\end{align}
where, in our conventions, 
\begin{align}
\overstar{R}_{\mu \nu \;\;\;\, \sigma}^{\;\;\;\;\;\lambda} =\tensor{e}{_a^\lambda}\tensor{e}{^b_\sigma}\overstar{R}_{\mu \nu \;\;\;\,b }^{\;\;\;\;\;a}= \partial_\mu \overstar{\Gamma}_{\nu \sigma}^{\; \; \; \; \lambda}-\partial_\nu \overstar{\Gamma}_{\mu \sigma}^{\; \; \; \; \lambda} + \overstar{\Gamma}_{\mu \rho}^{\; \; \; \; \lambda} \overstar{\Gamma}_{\nu \sigma}^{\; \; \; \; \rho}-\overstar{\Gamma}_{\nu \rho}^{\; \; \; \; \lambda} \overstar{\Gamma}_{\mu \sigma}^{\; \; \; \; \rho}.
\end{align}
By means of Eq. \eqref{Gamma-star}, we can write
\begin{align}
\overstar{R}_{\mu \nu \;\;\;\, \sigma}^{\;\;\;\;\;\lambda} &=\tensor{\cR}{_\mu_\nu^\l_\s}  + 3 G^{\epsilon \lambda} \left(\nabla^{(G)}_\mu \mathcal{K}_{[\nu \sigma \epsilon]}-\nabla^{(G)}_\nu \mathcal{K}_{[\mu \sigma \epsilon]}  \right) + 9 G^{\epsilon \lambda} G^{\alpha \rho} \left(\mathcal{K}_{[\mu \rho \epsilon]} \mathcal{K}_{[\nu \sigma \alpha]}-\mathcal{K}_{[\nu \rho \epsilon]} \mathcal{K}_{[\mu \sigma \alpha]}\right),
\end{align}
where $\tensor{\cR}{_\mu_\nu^\l_\s} $ is the Riemann  tensor for the Levi-Civita connection associated with the effective metric $G^{\mu\nu}$ (cf. Eq. \eqref{curly-Riemann-tensor}).  We note that the above equation shows that the relation between $\overstar{R}_{\mu \nu \;\;\;\, \sigma}^{\;\;\;\;\;\lambda}$ and $\tensor{\cR}{_\mu_\nu^\l_\s}$ has  the same functional form  as the formula relating the Riemann tensor of Einstein-Cartan theory to the Riemann tensor of general relativity (see e.g. Eq. (75) in Ref. \cite{Battista-DeFalco}).

It is important to stress that, despite  the presence of a Lorentz violating term in the Dirac equation \eqref{quantum-Dirac-equation}, our model predicts a spin precession equation \eqref{spin-precession} and translational motion  \eqref{acceleration} having the same form as in standard Einstein-Cartan theory (cf. Ref. \cite{Audretsch1981b}). Our analysis proves that this  is true in any background geometry, not only the particular FLRW model discussed in Sec. \ref{Sec:Cosmic-solution} below. In particular, an interesting consequence of Eq. \eqref{acceleration} is that it predicts a gyro-gravitational factor  equals to one (as  can be seen from the numerical factor $\frac{ \hbar}{2m}$ on the right hand side of Eq. \eqref{acceleration}), as in the ordinary gravity theories \cite{deOliveira1962}, where this result can be ascribed to the fact  that the spinor field describes  particles having equal gravitational and inertial masses. This assures that the intrinsic spin behaves as if the particle was a gyroscope. Standard results can be obtained also for the gyromagnetic factor, as will be pointed out in the next section.

\subsubsection{The magnetic dipole moment}\label{Sec:magnetic-dipole}

The knowledge of the magnetization current permits the evaluation of the magnetic dipole moment of the Dirac particle (see e.g. Ref. \cite{Sakurai-Advanced} for the analogous flat-space case). Let us indeed consider the \qm{magnetization piece} of the interaction Lagrangian
\begin{align}
\mathcal{L}_{\rm M}^{\rm int}= \sqrt{-G}\mathcal{J}^{\mu}_{\rm M} A_\mu,
\end{align}
where $A_\mu$ is an external electromagnetic field and hereafter we set the electric charge $e=1$; such a coupling  $\cJ^\mu A_\mu$ arises naturally in the appropriate setup, i.e. on a stack of branes in the IKKT model.
Bearing in mind Eq. \eqref{magnetization-current}, we can write 
\begin{align}
\mathcal{L}_{\rm M}^{\rm int}= \frac{i \hbar}{2m} \sqrt{-G} \left[\widehat{D}_\nu \left(A_\mu \frac{\obar{\Psi}\sigma^{\mu \nu} \Psi}{\rho}\right) - \frac{\obar{\Psi}\sigma^{\mu \nu} \Psi}{\rho} \widehat{D}_\nu A^\mu\right].
\label{L-int-M-2}
\end{align}
Since $\obar{\Psi}\sigma^{\mu \nu} \Psi$ behaves as a tensor under general coordinate transformations,  we can define the vector
\begin{align}
B^\nu := \frac{\obar{\Psi}\sigma^{\mu \nu} \Psi}{\rho} A_\mu,
\end{align}
and write Eq. \eqref{L-int-M-2} in terms of the torsion-free covariant derivative $\nabla^{(G)}_\nu$. In this way, we obtain the general expression
\begin{align}
\mathcal{L}_{\rm M}^{\rm int}&= \frac{i \hbar}{2m} \sqrt{-G} \left(\nabla^{(G)}_\nu B^\nu - \frac{\obar{\Psi}\sigma^{\mu \nu} \Psi}{\rho} \nabla^{(G)}_\nu A^\mu\right)
\nn \\ 
&= \frac{i \hbar}{2m} \partial_\nu \left(\sqrt{-G}B^\nu \right) + \sqrt{-G} \frac{i \hbar}{2m \rho} \obar{\Psi}\sigma^{\mu \nu} \Psi \left(\frac{1}{2}\overstar{F}_{\mu \nu}+ 3 \mathcal{K}_{[\mu \nu \epsilon]}A^\epsilon\right),
\label{L-int-M-3}
\end{align}
where we have exploited Eq. \eqref{Gamma-star} and 
\begin{align}
\overstar{F}_{\mu \nu} =2 \overstar{\nabla}_{[\mu}A_{\nu]} = 2 \nabla^{(G)}_{[\mu}A_{\nu]} - 6 \mathcal{K}_{[\mu \nu \epsilon]}A^\epsilon,  
\label{F-star-mu-nu}
\end{align}
is the electromagnetic field strength. 

At this stage, if we suppose that the vector $B^\nu$ falls off rapidly enough at infinity, then the total derivative occurring in Eq. \eqref{L-int-M-3} can be ignored. Let us also consider the geometric background provided by the FLRW cosmological solution, which will be introduced in the next section. In this geometry, we have  $\mathcal{K}_{[\mu \nu \epsilon]} =0$  (cf. Eq. \eqref{CONtorsion-BG-explicit} below). Then, Eq. \eqref{L-int-M-3} becomes
\begin{align}
\left. \mathcal{L}_{\rm M}^{\rm int}\right \vert_{\rm FLRW}= \sqrt{-G}\frac{i \hbar}{2m \rho} \obar{\Psi}\sigma^{\mu \nu} \Psi \left(\frac{1}{2}\widehat{F}_{\mu \nu}\right),
\end{align}
with $\widehat{F}_{\mu \nu}=2 \nabla^{(G)}_{[\mu}A_{\nu]}$ (see Eq. \eqref{F-star-mu-nu}).

Due to the conservation law \eqref{conservation-eff-J-mu}, the proper normalization of the fermions is obtained by absorbing the factor $\r^{-1}$ in the spinor:
\begin{align}
    \chi = \r^{-{1/2}}\, \Psi \ .
\end{align}
Then the interaction with the electromagnetic field takes the standard form 
\begin{align}
\left. \mathcal{L}_{\rm M}^{\rm int}\right \vert_{\rm FLRW}= \sqrt{-G}\frac{i \hbar}{2m} \obar{\chi}\sigma^{\mu \nu} \chi \left(\frac{1}{2}\widehat{F}_{\mu \nu}\right),
\end{align}
and hence, upon working out the nonrelativistic limit of the last formula, the magnetic dipole moment $\mu_{\rm D}$ of the Dirac particle  turns out to be (at tree level)
\begin{align}
\mu_{\rm D} = \frac{\hbar}{2 m},   
\end{align}
yielding for  the gyromagnetic ratio  the same value as in flat spacetime, i.e., 
\begin{align}
    g_{\rm D} = 2.
\end{align}

In the case of a generic background, the interaction Lagrangian \eqref{L-int-M-3} will include also the term  $\mathcal{K}_{[\mu \nu \epsilon]}$ and hence the magnetic dipole moment will be influenced by the contributions coming from the contorsion tensor. This could lead to intriguing implications which might also permit the detection of torsion effects. 

\section{A particular cosmological background solution} \label{Sec:Cosmic-solution}

The analysis of Sec. \ref{Sec:particle-limit-Dirac} applies to a generic curved background provided by the IKKT matrix model. In this section,  we consider a particular background 
solution $\cM^{3,1}$ of the matrix model which describes a cosmological FLRW spacetime \cite{Sperling:2019xar}. It is worth recalling that we have evaluated the propagation of a scalar field in this setup in Ref. \cite{Battista-Steinacker-2022}. 

The frame defined by the FLRW background is, in Cartesian coordinates $x^\mu$, (cf. Eqs. \eqref{frame-E-def} and \eqref{eff-frame-and-rho})
\begin{align}
\tensor{E}{_{a}^\mu} &= \left(\sinh \eta \right)  \delta^a_\mu \ ,
\label{effective-frame-expression}
\\
\tensor{\mathcal{E}}{^{a}_\mu} &= \r \tensor{E}{^{a}_\mu}.
\label{rescaled-frame-def}
\end{align}
Bearing in mind Eq. \eqref{eff-metric}, the effective metric $G_{\mu \nu}$ can be written in terms of the auxiliary metric 
\begin{align}
\gamma^{\mu \nu} = \left(\sinh^2 \eta \right) \eta^{\mu \nu},
\label{metric-gamma-and-eta}
\end{align}
 as
\begin{align}
 G_{\mu\nu} &= \rho^2 \, \gamma_{\mu\nu},
\label{eff-metric-G}
\end{align}
where 
\begin{align}
\rho^2= \vert  \sinh \eta \vert^{3},
\label{rho-squared}
\end{align}
represents the dilaton. The symplectic volume form  $\r_M d^4 y$  can be written, in Cartesian coordinates $x^\mu$, as (cf. Eq. \eqref{rho-def})  
\begin{align}
  \r_M = \frac{1}{|\sinh\eta|} \ .
\end{align}
Explicitly,
the $SO(3,1)$-invariant  FLRW effective metric reads \cite{Sperling:2019xar}
\begin{align}
 d s^2_G = G_{\mu\nu} d x^\mu d x^\nu 
   &= -R^2 \vert \sinh \eta \vert^3 d \eta^2 + R^2 \vert \sinh \eta \vert \cosh^2 \eta \, d \Sigma^2 \ = -d t^2 + a^2(t)d\Sigma^2,
   \label{eff-metric-FRW}
\end{align}
where  $a(t)$ is the cosmic  scale factor  and
\begin{align}
    d\Sigma^2 = d\chi^2 + \sinh^2\chi (d\theta^2 + \sin^2 \theta d\varphi^2),
    \label{dSigma2}
\end{align}
the invariant length element on the space-like hyperboloids $H^3$ (with $-\infty \leq \chi <\infty$, $0 \leq \theta < \pi$, $0 \leq \varphi <2\pi$).

The Weitzenb\"ock connection $\tensor{\Gamma}{_\nu_\rho^\mu}$ associated to the frame $\tensor{E}{_a ^\mu}$ is 
obtained from Eq. \eqref{effective-frame-expression} as
\begin{align}
\tensor{\Gamma}{_\nu_\lambda^\mu} &= -\tensor{E}{^a_\lambda} \del_\nu\tensor{E}{_a^\mu} 
 = -\frac{1}{\sinh \eta}\d^\mu_\lambda \del_\nu  \sinh\eta,  
 \label{Gamma-BG-explicit}
\end{align}
which, in turn, leads to
\begin{align}
    \tensor{\Gamma}{^\alpha_\lambda_\beta}= \gamma^{\alpha \nu} \gamma_{\beta \mu} \tensor{\Gamma}{_\nu_\lambda^\mu} 
    =\frac{1}{\rho^2 R^2} \tau^\alpha G_{\lambda \beta}.
\end{align}
Here we have exploited Eqs. \eqref{metric-gamma-and-eta}--\eqref{rho-squared}, and we have 
introduced the $SO(3,1)$-invariant cosmic timelike vector field $\tau =  a(t) \partial_t$, 
satisfying the relations \cite{Steinacker:2019dii,Steinacker(2020),Fredenhagen(2021)}
\begin{subequations}
\begin{align}
&\left( R^2\sinh\eta \right) \del_\mu \sinh\eta = -\eta_{\mu\nu}\t^\nu,
\label{sinh-derivative-BG}
\\
& G_{\mu \nu }\t^\mu\t^{\nu} = -R^2 \cosh^2 \eta  \left \vert \sinh \eta \right \vert= - a^2(t).
\label{tau-length}
\end{align}
\end{subequations}
The torsion and contorsion tensors of the
Weitzenb\"ock connection \eqref{Gamma-BG-explicit} are given by, respectively,
\begin{align}
\tensor{ T}{_{\r}_{\s}^\mu} &= \tensor{ \Gamma}{_{\r}_{\s}^\mu}-\tensor{ \Gamma}{_\s_\r^\mu}
=  \frac{1}{R^2\r^2}\big( \d_\s^{\mu} \t_\r - \d^\mu_\r \t_\s \big), 
 \label{torsion-BG-explicit}
 \\
  \tensor{ K}{_{\mu}_{\nu}^{\s}}&=\frac{1}{2} \left(\tensor{ T}{_{\mu}_{\nu}^{\s}}+\tensor{T}{^\sigma_\mu_\nu} -\tensor{T}{_\nu^\sigma_\mu}\right)
 = -\tensor{K}{_\mu^\sigma_\nu}=\frac 1{R^2\r^2} (  G_{\mu\nu} \t^\s  -  \d_\mu^{\s} \t_\nu ), 
 \label{CONtorsion-BG-explicit}
\end{align}
where $\t_\nu :=  G_{\nu\s}\t^\s$. 
Further details on the geometry of the cosmological background can be found in Appendix \ref{Appendix:formulae-cosmol-sol}. 

As a consequence of  Eq. \eqref{sinh-derivative-BG},    the Dirac equation   becomes (cf. Eq. \eqref{quantum-Dirac-equation})
\begin{align}
\gamma^\mu \widehat{D}_\mu \Psi + m \Psi + \frac{3}{4} \frac{\tau_\mu}{\rho^2 R^2} \gamma^\mu \Psi=0,  
\label{quantum-Dirac-equation-BG}
\end{align}
 where we have taken into account that in  the FLRW geometry we have   
 \begin{align}
\mathcal{K}_{[\alpha \beta \gamma]} =0,  
\label{total-antisymm-cont-FLRW}
 \end{align}
 owing to  Eq. \eqref{CONtorsion-BG-explicit} (see also Eq. \eqref{Levi-contorsion-full}). The term involving the cosmic vector field $\tau_\mu$ is responsible for the breaking of  the local Lorentz invariance, which can be attributed to the dilaton. It is thus clear that the investigation of Sec. \ref{Sec:particle-limit-Dirac} can be performed also within the geometrical setup \eqref{eff-metric-FRW}. However, in this case the analysis greatly simplifies due to Eq. \eqref{total-antisymm-cont-FLRW}. 

\section{Conclusions}\label{Sec:concluosions}

In this paper, we have examined the evolution of a Dirac particle on a generic curved 3+1-dimensional background brane within the IKKT matrix model.
This is non-trivial due to the non-standard form of the fermionic action and  the absence of manifest local Lorentz invariance.
We show that despite the different origin, 
the fermionic action differs from the one in general relativity only through a coupling to the totally antisymmetric part of the Weitzenb\"ock (con)torsion, which is determined by the effective frame. This extra term vanishes on a specific cosmological background  \cite{Sperling:2019xar}, where the propagation of scalar fields was studied in \cite{Battista-Steinacker-2022}.

We then examine the coupling of fermions in the 
present model in more detail by means of the JWKB approximation scheme. This permits to analyze  first-order non-trivial quantum corrections characterizing the dynamics of the fermion. Despite the different origin of the action, both the spin precession and the translation motion assume the same form as in Einstein-Cartan theory.
More specifically, we have shown that our  Eqs \eqref{spin-precession}  and  \eqref{acceleration} are analogous to the  equations of motion governing the dynamics of a Dirac fermion in a Riemann-Cartan spacetime. As a consequence, we find a gyro-gravitational factor in agreement with the predictions of standard gravity models. On the other hand, the gyromagnetic ratio assumes the usual (tree-level) value only if we consider the particular case of the FLRW background geometry \eqref{eff-metric-FRW}, whereas in the most general situations it receives contributions originating from the (totally antisymmetric part of the)  contorsion tensor of the Weitzenb\"ock connection. This should lead to observable  physical consequences on non-trivial backgrounds, which in principle could be  tested experimentally. We leave a detailed assessment of such effects to future work.

\section*{Acknowledgement}

This work  is supported by the Austrian Science Fund (FWF) grant P32086.

\appendix
\numberwithin{equation}{section}

\section{Our conventions for the Dirac  matrices}\label{Appendix:Dirac-matrices-conventions}

We employ the following conventions for the Dirac matrices:
\begin{align}
\gamma^a &= \tensor{\mathcal{E}}{^a_\mu} \gamma^\mu,
\nn \\
\{\gamma^a,\gamma^b\}&= 2 \eta^{ab}\, \one,
\nn \\
\{\gamma^\mu,\gamma^\nu\} &= 2 G^{\mu \nu}\, \one \ ,
\nn \\
 \gamma^{a \, \dagger}  &= \gamma^{\widehat{0}} \gamma^a \gamma^{\widehat{0}},
\nn \\
(\gamma^5)^2 &= \one,
\nn \\
\{\gamma^5,\gamma^a\}&=0=\{\gamma^5,\gamma^\mu\},
\end{align}
where the matrices $\gamma^a$ read as
\begin{align}
\gamma^{\widehat{0}} = -i 
\begin{pmatrix}
\one & 0 \\
0 & -\one  
\end{pmatrix}, \qquad \gamma^{\widehat{a}} = -i \begin{pmatrix}
0 & \sigma^{\widehat{a}} \\
-\sigma^{\widehat{a}} & 0 
\end{pmatrix},    
\end{align}
the Pauli matrices being
\begin{align}
    \sigma^{\widehat{1}} = \begin{pmatrix}
    0 & 1 \\
1 & 0  
\end{pmatrix}, \quad
  \sigma^{\widehat{2}} = \begin{pmatrix}
    0 & -i \\
i & 0  
\end{pmatrix},
\quad  
\sigma^{\widehat{3}} = \begin{pmatrix}
    1 & 0 \\
0 & -1  
\end{pmatrix}.
\end{align}
Moreover, the fifth (flat) Dirac matrix reads as
\begin{align}
    \gamma^5 = \begin{pmatrix}
    0 & \one \\
    \one & 0
    \end{pmatrix}.
\end{align}
Last, it is worth noting that with our conventions we have
 \begin{align}
\left(\obar \Psi \gamma^a \Psi \right)^\dagger &= \obar \Psi \gamma^a \Psi, 
\nn \\
\left(i \obar \Psi  \Psi \right)^\dagger &= i \obar \Psi  \Psi,
\nn \\
\left(i \obar \Psi  \sigma^{ab} \Psi \right)^\dagger &=i \obar \Psi  \sigma^{ab} \Psi,
\end{align}
where we recall $\obar \Psi = \Psi^\dagger \gamma^{\widehat{0}}$ and $\sigma^{ab} = i \gamma^{[a}\gamma^{b]}$.

\section{Mass term from fuzzy extra dimensions}\label{Sec:mass-term}

We briefly discuss the origin of mass terms for fermions in the IKKT model.
Since the fermions in this model are (matrix-valued) Majorana-Weyl spinors of $SO(9,1)$, no mass term for the fermions is allowed in the action. However, fermions may acquire a mass through the Higgs effect in a non-trivial vacuum. This is where the 6 \qm{transversal} matrices $T^{A+3}, \ A = 4,...,9$ enter the stage: they play the role of scalar fields
\begin{align}
    \phi^A := T^{A+3}, \qquad  A=1,...,6  \ 
\end{align}
in the (equivalent) formulation of the IKKT model as noncommutative $\cN=4$ SYM on $\cM^{3,1}$  (note that this is a statement for the action; the background $\cM^{3,1}$ need not be supersymmetric or BPS) \cite{Aoki:1999vr}. 
Now assume that these scalar fields acquire non-trivial VEV's  
\begin{align}
    \langle\phi^A\rangle := K^A \neq 0, \qquad A=1,...,6 \ ,
    \label{VEV-scalars}
\end{align}
 for
 $K^A$ being generators of some fuzzy space $\cK$, such as fuzzy $S^2_N$ of fuzzy $\C P^2_N$.
At the classical level, this can be achieved e.g. by adding a suitable cubic term to the potential, cf. Refs. \cite{Aschieri:2006uw,Steinacker:2014lma}. At the quantum level, such backgrounds might arise even without adding such terms by hand. Assuming such a vacuum, the Yukawa couplings of $K^A$ contained in the fermionic action \eq{fermionic-action-geom} lead to terms   of the form 
\begin{align}
S = \Tr  \obar\Psi  \gamma_A [T^A,\Psi]
 \,\, \sim\,\, \int d^4 y\, \rho_M(y)\, \obar\Psi i \Delta_A [K^A,\Psi] \ .
 \label{Yukawa}
\end{align}
We can then expand the fermions -- and any other fields -- in terms of harmonics on $\cK$.
Rewritten in terms of 
the 3+1-dimensional fermions, it amounts to a mass term coupling the 4 Weyl or Majorana fermions in $\cN=4$ SYM; for details we refer to the literature \cite{Chatzistavrakidis:2009ix,Steinacker:2014lma}. 
This is what we have assumed in this paper, where we have proceeded with the analysis of 
3+1-dimensional fermions governed by the Lagrangian \eq{symmetrized-Lagrangian} 
in the presence of an extra mass term.

\section{More on the cosmological solution} \label{Appendix:formulae-cosmol-sol}

In this appendix, we give some further details regarding the cosmological background solution \eqref{eff-metric-FRW}.

The torsion and contorsion tensors of the Weitzenb\"ock connection \eqref{Gamma-BG-explicit} have been given in Eqs. \eqref{torsion-BG-explicit} and \eqref{CONtorsion-BG-explicit}, respectively. They satisfy the following useful relations \cite{Steinacker(2020)}:
\begin{align}
\tensor{ T}{_\r^{\mu}_\s}\tensor{ T}{_\nu_\mu^\r} 
&= \frac 1{R^{4}\r^4}( - \t_\s  \t_\nu + G_{\s\nu}  \t^\mu \t_\mu) = \tensor{ T}{_\r^{\mu}_\nu}\tensor{ T}{_\s_\mu^\r}, 
\nn\\ 
\tensor{ K}{_\mu^\r_\nu}\tensor{ K}{_\r^\mu_\s}
&= \frac 3{R^{4}\r^4} \t_\nu \t_\s, 
\nn\\
-\frac 12(\tensor{ T}{_\r^{\mu}_\s}\tensor{ T}{_\nu_\mu^\r} 
+ \tensor{ T}{_\r^{\mu}_\nu}\tensor{ T}{_\s_\mu^\r})
- \tensor{ K}{_\mu^\r_\nu}\tensor{ K}{_\r^\mu_\s}
&= -\frac 1{R^{4}\r^4}(2\t_\s  \t_\nu +  G_{\s\nu}  \t^\mu \t_\mu), 
\nn\\
2 \r^{-2} \del_\s \r \del_\nu \r &=  \frac{9}{2R^4\r^4} \t_\s \t_\nu,  
\nn\\
\r\Box_G\r &= \frac{3}{2R^2}\left(4 + \frac 12 \frac{\cosh^2 \eta}{\sinh^2\eta}\right),
\end{align}
where  $\Box_G \rho = - |G|^{-1/2}\,
    \del_\mu\big( |G|^{1/2}\,G^{\mu\nu}\del_\nu\rho\big)$ and we recall $\tau_\nu :=G_{\nu \sigma} \tau^\sigma$.

The Riemann  and  Ricci tensors for the Levi-Civita connection associated with the effective metric $G^{\mu\nu}$ are defined,  in terms of Christoffel symbols $\tensor{\G}{^{(G)}_\nu_\r^\l}$,  as
\begin{align}
\tensor{\cR}{_\mu_\nu^\l_\s} 
  &= \del_\mu \tensor{\G}{^{(G)}_\nu_\s^\l} - \del_\nu \tensor{\G}{^{(G)}_\mu_\s^\l}
 + \tensor{\G}{^{(G)}_\mu_\r^\l} \tensor{\G}{^{(G)}_\nu_\s^\r}  
 - \tensor{\G}{^{(G)}_\nu_\r^\l} \tensor{\G}{^{(G)}_\mu_\s^\r},  
\label{curly-Riemann-tensor}
\\
 \cR_{\nu\s} &=\tensor{\cR}{_\mu_\nu^\mu_\s} =\del_\mu \tensor{\G}{^{(G)}_\nu_\s^\mu} - \del_\nu \tensor{\G}{^{(G)}_\mu_\s^\mu}
 + \tensor{\G}{^{(G)}_\mu_\r^\mu} \tensor{\G}{^{(G)}_\nu_\s^\r}  
 - \tensor{\G}{^{(G)}_\nu_\r^\mu} \tensor{\G}{^{(G)}_\mu_\s^\r},
 \label{Riemann-tensor-gen}
\end{align}
respectively. In Riemann normal coordinates at $p\in\cM$, this simplifies using \eq{LC-contorsion-eff} as 
\begin{align}
\tensor{\cR}{_\mu_\nu^\l_\s} 
 &= \del_\mu (\tensor{\widetilde\G}{_\nu_\s^\l} - \tensor{\cK}{_\nu_\s^\l}) 
   - \del_\nu (\tensor{\widetilde\G}{_\mu_\s^\l} - \tensor{\cK}{_\mu_\s^\l}), 
   \nn\\
 \cR_{\nu\s} &= \del_\mu (\tensor{\widetilde\G}{_\nu_\s^\mu} - \tensor{\cK}{_\nu_\s^\mu}) 
              - \del_\nu (\tensor{\widetilde\G}{_\mu_\s^\mu} - \tensor{\cK}{_\mu_\s^\mu}) \ .
\end{align}
Now we exploit the fact that the curvature of the Weitzenb\"ock connection vanishes,
\begin{align}
 0
 &= \del_\mu \tensor{\widetilde\G}{_\nu_\s^\l} - \del_\nu\tensor{\widetilde\G}{_\mu_\s^\l}
 +  \tensor{\widetilde\G}{_\mu_\r^\l}  \tensor{\widetilde\G}{_\nu_\s^\r}  
 -  \tensor{\widetilde\G}{_\nu_\r^\l}  \tensor{\widetilde\G}{_\mu_\s^\r},  \nn\\
 0 &=\del_\mu  \tensor{\widetilde\G}{_\nu_\s^\mu} - \del_\nu  \tensor{\widetilde\G}{_\mu_\s^\mu}
 +  \tensor{\widetilde\G}{_\mu_\r^\mu}  \tensor{\widetilde\G}{_\nu_\s^\r}  
 -  \tensor{\widetilde\G}{_\nu_\r^\mu}  \tensor{\widetilde\G}{_\mu_\s^\r} , 
\end{align}
and  obtain the tensorial equations
\begin{align}
\tensor{\cR}{_\mu_\nu^\l_\s} 
 &= -\nabla^{(G)}_\mu \tensor{\cK}{_\nu_\s^\l}
   + \nabla^{(G)}_\nu  \tensor{\cK}{_\mu_\s^\l} 
   - \tensor{\cK}{_\mu_\r^\l} \tensor{\cK}{_\nu_\s^\r}  
   + \tensor{\cK}{_\nu_\r^\l} \tensor{\cK}{_\mu_\s^\r}, \nn\\
 \cR_{\nu\s} &= -\nabla^{(G)}_\mu \tensor{\cK}{_\nu_\s^\mu}
                + \nabla^{(G)}_\nu \tensor{\cK}{_\mu_\s^\mu}  
        - \tensor{\cK}{_\mu_\r^\mu} \tensor{\cK}{_\nu_\s^\r}  
        + \tensor{\cK}{_\nu_\r^\mu} \tensor{\cK}{_\mu_\s^\r},       
  \label{riemann-K}
\end{align}
 using 
  \begin{align}
 \qquad \tensor{\cK}{_{\mu}_{\nu}^{\s}} = \tensor{\widetilde\G}{_{\mu}_{\nu}^{\s}} \qquad \mbox{at} \ \ p \ .
\label{Weitzenbock-Torsion-RiemanNC}
\end{align}
In particular, the explicit expression of the Ricci tensor reads as
\begin{align}
 \cR_{ \nu \sigma} &=     \frac{5}{2}\frac{1}{\r^{4}R^4}  \t_\nu \t_\s + \frac{1}{2\r^2R^2} G_{ \nu \sigma} \left(6 - \coth^2 \eta\right). 
\end{align}
For further details we refer the reader to  Ref. \cite{Steinacker(2020)}.

\bibliography{references}

\end{document}